\documentclass[aps,preprint,prd,showpacs,nofootinbib]{revtex4}

\usepackage{amsmath,amssymb}
\usepackage{graphicx,subfigure}
\usepackage{color,multirow}
\usepackage[colorlinks,linkcolor=magenta,anchorcolor=cyan,citecolor=blue,plainpages=false]{hyperref}

\hypersetup{colorlinks=true,
    breaklinks=true,
    pdfstartview=Fit,
    linkcolor=blue,
    citecolor=blue,
    urlcolor=blue}

\def\be{\begin{equation}}
    \def\ee{\end{equation}}
\def\ba{\begin{eqnarray}}
    \def\ea{\end{eqnarray}}

\begin{document}

\title{Hint of $r\simeq 0.01$ after DESI DR2 ?}

\author{Hao Wang$^{1,2} $\footnote{\href{wanghao187@mails.ucas.ac.cn}{wanghao187@mails.ucas.ac.cn}}}
\author{Ze-Yu Peng$^{2,3} $ \footnote{\href{pengzeyv@mail.ustc.edu.cn}{pengzeyv@mail.ustc.edu.cn}}}
\author{Yun-Song Piao$^{1,2,4,5} $ \footnote{\href{yspiao@ucas.ac.cn}{yspiao@ucas.ac.cn}}}

    \affiliation{$^1$ School of Fundamental Physics and Mathematical
        Sciences, Hangzhou Institute for Advanced Study, UCAS, Hangzhou
        310024, China}

    \affiliation{$^2$ School of Physical Sciences, University of
        Chinese Academy of Sciences, Beijing 100049, China}

    \affiliation{$^3$ International Center for Theoretical Physics
        Asia-Pacific, Beijing/Hangzhou, China}

    \affiliation{$^4$ Institute of Theoretical Physics, Chinese
        Academy of Sciences, P.O. Box 2735, Beijing 100190, China}

    %   \date{}
    \begin{abstract}

In the report by BICEP/Keck collaborations, the tensor-to-scalar
ratio is $r_{0.05}<0.036$ (95\% C.L.). However, recent datasets
have preferred the evolving dark energy, which thus have
significantly shifted the bestfit values of standard $\Lambda$CDM
cosmological parameters. In this paper, we perform the joint
analysis of BICEP/Keck cosmic microwave background (CMB) B-mode
data, latest DESI DR2 baryon acoustic oscillations and supernova
data, combined with Planck PR3 and PR4 CMB data respectively, and
find $r_{0.05}=0.0159^{+0.0057}_{-0.014}$ and
$r_{0.05}=0.0164^{+0.0063}_{-0.014}$. The constraints on $r$ are
further tightened compared to the result of BICEP/Keck
collaborations. Though there might be still systematic
uncertainties in B-mode measurements due to the foreground
contamination, our work is to not say what the value of $r$ is,
but present the state-of-the-art constraints on $r$ and emphasize
that the detection for $r$ depends potentially on our insight into
the dark universe, highlighting the important role of cosmological
surveys in comprehending our very early universe.

    \end{abstract}

    \maketitle
    %\tableofcontents
    %\newpage

\section{INTRODUCTION}

The current paradigm of very early universe, inflation
\cite{Guth:1980zm,Linde:1981mu,Albrecht:1982wi,Starobinsky:1980te},
predicts a nearly scale-invariant scalar perturbation consistent
with recent observations, as well as the primordial gravitational
waves (GWs).

It is well known that the ultra-low-frequency primordial GWs at
$f\sim 10^{-18}-10^{-16}$Hz, which can source the B-mode
polarization in the cosmic microwave background (CMB)
\cite{Seljak:1996ti,Kamionkowski:1996zd,Seljak:1996gy}, is the
``smoking gun" of inflation. Based on the standard $\Lambda$CDM
model, using Planck18 CMB, pre-DESI baryon acoustic oscillations
(BAO) and BICEP/Keck18 CMB B-mode datasets the BICEP/Keck
collaboration has reported the tensor-to-scalar ratio \be
r_{0.05}=0.014^{+0.010}_{-0.011}\quad (68\%\,\,\mathrm{CL})\ee
with $r_{0.05}<0.036$ at 95\% C.L. \cite{BICEP:2021xfz}, also
\cite{Tristram:2021tvh}\footnote{This upper bound can be tighter
if the pre-recombination resolutions of the Hubble tension were
considered \cite{Ye:2022afu,Jiang:2023bsz}.}.

Though the standard $\Lambda$CDM model is thought to be the most
successful model explaining most of cosmological observations,
recently using their first year data the DESI collaboration
\cite{DESI:2024mwx,DESI:2024aqx,DESI:2024kob} has found that DE is
evolving at $\gtrsim 3\sigma$ significance level. This result will
inevitably bring the shifts of the bestfit values of relevant
$\Lambda$CDM cosmological parameters and possibly alter the
amplitude of lensing B-mode spectrum, so that $r\simeq 0.01$ might
emerge more significantly \cite{Wang:2024sgo,Wang:2024tjd}. Though
the scientific community still have doubts about systematic errors
of DESI, latest DESI DR2 \cite{DESI:2025zgx} are consistent with
DESI DR1, and thus further strengthened the results with DESI DR1
\cite{Lodha:2025qbg}\footnote{The relevant issues have been also
intensively investigated since DESI DR1 and DR2,
e.g.\cite{Luongo:2024fww,Cortes:2024lgw,Carloni:2024zpl,Colgain:2024xqj,Giare:2024smz,Wang:2024dka,Park:2024jns,Wang:2024pui,Shlivko:2024llw,Dinda:2024kjf,Seto:2024cgo,Bhattacharya:2024hep,Roy:2024kni,Wang:2024hwd,Notari:2024rti,Heckman:2024apk,Gialamas:2024lyw,Orchard:2024bve,Colgain:2024ksa,Li:2024qso,Ye:2024ywg,Giare:2024gpk,Dinda:2024ktd,Jiang:2024viw,Alfano:2024jqn,Jiang:2024xnu,Sharma:2024mtq,Ghosh:2024kyd,Reboucas:2024smm,Pang:2024qyh,Wolf:2024eph,RoyChoudhury:2024wri,Arjona:2024dsr,Wolf:2024stt,Giare:2024ocw,Alestas:2024eic,Carloni:2024rrk,Bhattacharya:2024kxp,Specogna:2024euz,Li:2024qus,Ye:2024zpk,Pang:2024wul,Akthar:2024tua,Colgain:2024mtg,daCosta:2024grm,Chan-GyungPark:2025cri,Sabogal:2025mkp,Du:2025iow,Ferrari:2025egk,Jiang:2025ylr,Peng:2025nez,Jiang:2025hco,Wolf:2025jlc,Giare:2025pzu,Feng:2025mlo,Hossain:2025grx,Chakraborty:2025syu,Borghetto:2025jrk,Pan:2025psn,Pang:2025lvh,Wolf:2023uno}}.

%Thus it is necessary and significant to investigate the underlying
%impact of DESI DR2 on the search for $r$.

\begin{figure*}
    \includegraphics[width=0.8\columnwidth]{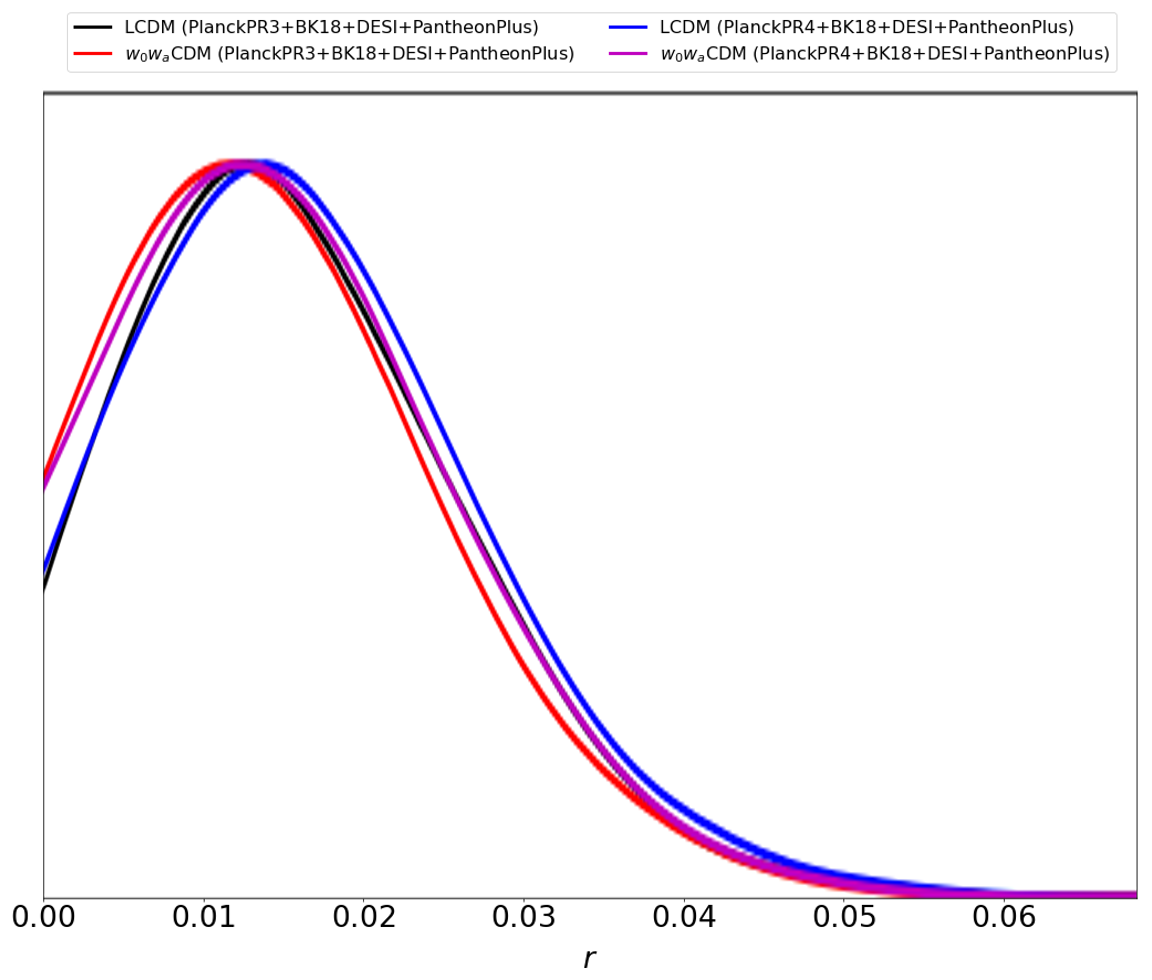}
\caption{\label{nsr} The 1D posteriors of $r$ for different models
and datasets.}
\end{figure*}

Here, we focus on the impact of DESI DR2 for the standard
cosmological model and present the state-of-the-art constraints on
$r$, \be r_{0.05}=0.0159^{+0.0057}_{-0.014}\quad
(68\%\,\,\mathrm{CL})\ee \be
r_{0.05}=0.0164^{+0.0063}_{-0.014}\quad (68\%\,\,\mathrm{CL})\ee
for the datasets with Planck PR3 and Planck PR4+ACT lensing data,
respectively.
%and \be
%r_{0.05}=0.0160^{+0.0060}_{-0.013}\quad (68\%\,\,\mathrm{CL})\ee
%for the datasets with DESY5.
The $1\sigma$ upper bound is more tightened than that reported by
the BICEP/Keck collaboration \cite{BICEP:2021xfz}, and the bestfit
$r$ is $r_{0.05}\sim 0.01$ with different values depending on
models and datasets, see also Fig.\ref{nsr}. Our results indicate
that the detection for $r$ can depend potentially on our insight
into the nature of DE, highlighting the important role of
cosmological surveys in comprehending our very early universe.

%\begin{figure*}
%    \includegraphics[width=\columnwidth]{nsrsn.png}
%\caption{\label{nsrsn} The 1D posteriors of $r$ and $n_s$ for the
%$w_0w_a$CDM model, where $n_s$ is the spectral index of primordial
%scalar perturbation.}
%\end{figure*}

\section{Datasets and Results}

Here, we use \textbf{DESI DR2} BAO data. Recent \textbf{DESI DR2}
consists of bright galaxies, LRGs, ELGs, quasars and Ly$\alpha$
Forest samples at the redshift region $0.1<z<4.2$
\cite{DESI:2025zgx}, which is consistent with SDSS and DESI DR1
\cite{DESI:2024mwx}.

%their measurements for the comoving distances $D_M(z)/r_d$ and
%$D_H(z)/r_d$,
% as listed in Table \ref{DESI},
%where
%    \begin{equation}\label{DMDH}
%        D_M(z)\equiv\int_{0}^{z}{cdz'\over H(z')},\quad D_H(z)\equiv {c\over
%        H(z)},
%    \end{equation} and $r_d=\int_{z_d}^{\infty}{c_s(z)\over
%H(z)}$ is the sound horizon with $z_d\simeq1060$ at the baryon
%drag epoch and $c_s$ the speed of sound, as well as the
%angle-averaged quantity $D_V/r_d$, where
%$D_V(z)\equiv\left(zD_M(z)^2D_H(z)\right)^{1/3}$.

To perform the search for $r$, we use the CMB B-mode \textbf{BK18}
data \cite{BICEP:2021xfz} combined with \textbf{Planck PR3}
dataset (low-T \texttt{Commander}, low-E \texttt{SimALL} and
Planck 2018 high-$l$ TT, TE, EE spectra, and reconstructed CMB
lensing spectrum
\cite{Planck:2018vyg,Planck:2019nip,Planck:2018lbu}) and
\textbf{Planck PR4} (low-T \texttt{Commander}, low-E
\texttt{SimALL} and \texttt{Camspec} high-$l$
likelihood\cite{Rosenberg:2022sdy}, as well as Planck PR4
lensing\cite{Carron:2022eyg}), respectively. In addition, we also
use the supernova dataset: \textbf{PantheonPlus} (consisting of
1701 light curves of 1550 spectroscopically confirmed Type Ia SN
coming from 18 different surveys \cite{Scolnic:2021amr}) and
\textbf{DES-Y5} (Dark Energy Survey, as part of their Year 5 data,
recently published results based on a new, homogeneously selected
sample of 1635 photometrically classified SN Ia with redshifts
$0.1<z<1.3$, which is complemented by 194 low-redshift SN Ia in
common with the Pantheon+ sample spanning $0.025<z<0.1$
\cite{DES:2024jxu}).

%Throughout this paper we work with
%\textbf{Planck18+BK18+DESI+PantheonPlus} datasets.

%We also include additional \textbf{SH0ES} Cepheid distances as a
%calibrator of the SN1a magnitude for comparison
%\cite{Riess:2021jrx}.

%\begin{table}[htbp]
%    \centering
%    \begin{tabular}{c|c|cc|c|c}
%        Tracer&$z_\mathrm{eff}$&$D_M/r_d$&$D_H/r_d$&$D_V/r_d$&$r_{M,H}$\\
%        \hline
%        BGS&0.295&-&-&$7.944\pm0.075$&-\\
%        LRG1&0.510&$13.587\pm0.169$&$21.863\pm0.427$&$12.720\pm0.098$&-0.475\\
%        LRG2&0.706&$17.347\pm0.180$&$19.458\pm0.332$&$16.048\pm0.110$&-0.423\\
%        LRG3+ELG1&0.934&$21.574\pm0.153$&$17.641\pm0.193$&$19.720\pm0.091$&-0.425\\
%        ELG2&1.321&$27.605\pm0.320$&$14.178\pm0.217$&$24.256\pm0.174$&-0.437\\
%        QSO&1.484&$30.519\pm0.758$&$12.816\pm0.513$&$26.059\pm0.400$&-0.489\\
%        Lya QSO&2.330&$38.988\pm0.531$&$8.632\pm0.101$&$31.267\pm0.256$&-0.431\\
%    \end{tabular}
%    \caption{\label{DESI}Statistics for the DESI samples of the DESI
%        DR2 BAO measurements used in this paper.}
%\end{table}

In our analysis, we consider the $w_0w_a$CDM model (well-known CPL
parameterisation for DE) where the state equation of DE is
\cite{Chevallier:2000qy,Linder:2002et} \be
w_\mathrm{DE}=w_0+w_a{z\over 1+z}.\ee Here, our MCMC analysis is
performed by using the \textbf{MontePython-3.6} sampler
\cite{Audren:2012wb,Brinckmann:2018cvx} and \textbf{CLASS} codes
\cite{Lesgourgues:2011re,Blas:2011rf}. The threshold of
Gelman-Rubin convergence criterion is $R-1<0.01$. The
corresponding priors of MCMC parameters are listed in
Table.\ref{prior}, and the pivot scale of $r$ is set to
0.05(Mpc)$^{-1}$.

    \begin{table}[htbp]
    \centering
    \begin{tabular}{cc}
        \hline
        Parameters&Prior\\
        \hline
        $100\omega_b$&[None, None]\\
        $\omega_{cdm}$&[None, None]\\
        $H_0$&[65, 80]\\
        $\ln10^{10}A_s$&[None, None]\\
        $n_s$&[None,None]\\
        $\tau_{reio}$&[0.004, None]\\
        \hline
        $w_0$&[-2, -0.34]\\
        $w_a$&[-3,2]\\
        \hline
        $r$&[0,0.5]\\
        \hline
    \end{tabular}
\caption{\label{prior} The priors of parameters we adopt in MCMC
analysis.}
 \end{table}

%\section{Results}

In Table \ref{mctab}, we present our MCMC results. In $w_0w_a$CDM
model with \textbf{Planck PR4}, an evolving DE is preferred with
$w_0=-0.841\pm0.054$ and $w_a=-0.595\pm0.202$, consistent with
those of DESI collaboration\cite{DESI:2025zgx}. The result of
tensor-to-scalar ratio $r$ is $r_{0.05}=0.0164^{+0.0063}_{-0.014}$
at the 1$\sigma$ (68\%) CL.. In particular, the upper bound of $r$
is suppressed in $w_0w_a$CDM (with smaller $n_s$ and $H_0$) than
that in $\Lambda$CDM, with $r_{0.05}<0.034$ and $r_{0.05}<0.035$
(95\% C.L.) respectively combined with both \textbf{Planck PR3}
and \textbf{Planck PR4}, as displayed in Fig.\ref{nsr}. In
addition, both $\Lambda$CDM and $w_0w_a$CDM with \textbf{Planck
PR4} prefer slightly smaller $n_s$ and $A_s$, as well as a
slightly larger bestfit $r$ than those with \textbf{Planck PR3}.

%The constraint on $r$ is slightly tighter than that with DESI DR1
%\cite{Wang:2024sgo}.

\begin{table*}[htbp]
    \centering
    \begin{tabular}{c|c|c|c|c}
        \hline
        \multirow{3}{*}{Parameters}&\multicolumn{4}{c}{\textbf{BK18+DESI+PantheonPlus}}\\
        \cline{2-5}
        &\multicolumn{2}{c|}{\textbf{+Planck PR3}}&\multicolumn{2}{c}{\textbf{+Planck PR4}}\\
        \cline{2-5}
        &$\Lambda$CDM&$w_0w_a$CDM&$\Lambda$CDM&$w_0w_a$CDM\\
        \hline
        $100\omega_b$&2.244(2.232)$\pm$0.013&2.236(2.237)$\pm$0.014&2.231(2.232)$\pm$0.012&2.223(2.215)$\pm$0.013\\
        $\omega_{cdm}$&0.118(0.118)$\pm$0.001&0.119(0.119)$\pm$0.001&0.118(0.118)$\pm$0.001&0.119(0.119)$\pm$0.001\\
        $H_0$&68.23(67.88)$\pm$0.30&67.54(67.93)$\pm$0.62&68.11(68.14)$\pm$0.28&67.50(67.41)$\pm$0.60\\
        $\ln10^{10}A_s$&3.049(3.038)$\pm$0.015&3.041(3.041)$\pm$0.014&3.048(3.037)$\pm$0.014&3.038(3.044)$\pm$0.014\\
        $n_s$&0.970(0.968)$\pm$0.003&0.967(0.968)$\pm$0.004&0.968(0.967)$\pm$0.003&0.965(0.966)$\pm$0.004\\
        $\tau_{reio}$&0.060(0.055)$\pm$0.007&0.055(0.052)$\pm$0.007&0.059(0.054)$\pm$0.007&0.054(0.057)$\pm$0.007\\
        \hline
        $w_0$&-&-0.834(-0.899)$\pm$0.055&-&-0.841(-0.840)$\pm$0.054\\
        $w_a$&-&-0.619(-0.416)$\pm$0.205&-&-0.595(-0.583)$\pm$0.202\\
        $r$&$0.0168(0.123)^{+0.0066}_{-0.013}$&0.0159(0.0066)$^{+0.0057}_{-0.014}$&0.0174(0.0107)$^{+0.0069}_{-0.014}$&$0.0164(0.0175)^{+0.0063}_{-0.014}$\\
        \hline
        $\Omega_m$&0.303(0.307)$\pm$0.004&0.312(0.308)$\pm$0.006&0.303(0.303)$\pm$0.004&0.311(0.312)$\pm$0.006\\
        $S_8$&0.811(0.813)$\pm$0.008&0.825(0.824)$\pm$0.009&0.810(0.805)$\pm$0.008&0.822(0.825)$\pm$0.009\\
        \hline
    \end{tabular}
    \caption{\label{mctab}Mean (bestfit) values and 1$\sigma$ regions
of the parameters of $\Lambda$CDM and $w_0w_a$CDM models. The
datasets are \textbf{Planck PR3+BK18+DESI+PantheonPlus} and
\textbf{Planck PR4+BK18+DESI+PantheonPlus}, respectively.}
\end{table*}
%\begin{figure*}
%    \includegraphics[width=\columnwidth]{mc.png}
%    \caption{\label{mc}2D contours at 68\% and 95\% CL for
%        the parameters of the $w_0w_a$CDM model fitting to \textbf{Planck18+BK18+DESI+PantheonPlus+SH0ES} and \textbf{Planck18+BK18+DESI+PantheonPlus}
%        datasets respectively.}
%\end{figure*}

In Table.\ref{mcsn}, we list our results for different SN data. When \textbf{SH0ES}
Cepheid-calibrated SN1a magnitude (equivalently SH0ES prior
\cite{Riess:2021jrx}) is considered, the tensor-to-scalar ratio
$r$ is similar but with larger $n_s$ and $H_0$.
%The $w_0-w_a$ contour shifts more to phantom-like $w_0+w_a<-1$.
%$S_8$ seems to also be slightly decreased with a larger $H_0$ and
%a smaller $\Omega_m$.
In $w_0w_a$CDM model with \textbf{DESY5} SN data
%to investigate the
%impact of different SN data on the constraint of $r$. The results
$r=0.0160^{+0.0060}_{-0.013}$, which has a larger lower bound than
that with PantheonPlus. However, replacing PantheonPlus with DESY5
also slightly lowers $H_0$ and exacerbates the Hubble tension,
which remains to be further investigated.

\begin{table*}[htbp]
    \centering
    \begin{tabular}{c|c|c}
        \hline
        \multirow{2}{*}{Parameters}&\multicolumn{2}{c}{\textbf{Planck PR3+BK18+DESI}}\\
        \cline{2-3}
        &\textbf{+PantheonPlus+SH0ES}&\textbf{+DESY5}\\
        \hline
        $100\omega_b$&2.239(2.233)$\pm$0.014&2.233(2.232)$\pm$0.014\\
        $\omega_{cdm}$&0.119(0.119)$\pm$0.001&0.119(0.120)$\pm$0.001\\
        $H_0$&69.18(69.11)$\pm$0.56&66.82(66.82)$\pm$0.56\\
        $\ln10^{10}A_s$&3.040(3.026)$\pm$0.014&3.040(3.050)$\pm$0.015\\
        $n_s$&0.966(0.967)$\pm$0.004&0.966(0.965)$\pm$0.004\\
        $\tau_{reio}$&0.054(0.047)$\pm$0.007&0.054(0.059)$\pm$0.008\\
        \hline
        $w_0$&-0.895(-0.895)$\pm$0.055&-0.753(-0.730)$\pm$0.057\\
        $w_a$&-0.611(-0.627)$\pm$0.218&-0.854(-0.970)$\pm$0.222\\
        $r$&0.0159(0.0217)$^{+0.0058}_{-0.014}$&$0.0160(0.0123)^{+0.0060}_{-0.013}$\\
        \hline
        $\Omega_m$&0.297(0.298)$\pm$0.005&0.319(0.320)$\pm$0.006\\
        $S_8$&0.821(0.818)$\pm$0.009&0.830(0.839)$\pm$0.009\\
        \hline
    \end{tabular}
    \caption{\label{mcsn}Mean (bestfit) values and 1$\sigma$ regions
        of the parameters of the $w_0w_a$CDM model. The datasets are
        \textbf{Planck PR3+BK18+DESI+PantheonPlus+SH0ES} and
        \textbf{Planck PR3+BK18+DESI+DESY5}, respectively.}
\end{table*}

%In Table. \ref{mctab} $w_0$ and $w_a$ are more different than
%other parameters fitting to BK18 data. Consistent with the
%previous work the evolving DE supported by DESI BAO data has a
%significant impact on $A_s$, $\Omega_m$ and thus
%$r$\cite{Wang:2024sgo}.

%Most part of $w_0-w_a$
%contour accounts for a larger $C_{l,\mathrm{lensing}}^{BB}$ with
%\textbf{SH0ES} and thus gives a smaller $r$ than fitting to
%\textbf{Planck18+BK18+DESI+PantheonPlus}.

%\begin{figure*}
%    \includegraphics[width=0.6\columnwidth]{lens.png}
%    \caption{\label{lens} 2D contours for $w_0-w_a$ fitting to \textbf{Planck18+BK18+DESI+PantheonPlus} and \textbf{Planck18+BK18+DESI+PantheonPlus+SH0ES} datasets. We compare $C_{l,\mathrm{lensing}}^{BB}$ at $l=100$ where is constrained by BK18 data as shown in Fig.\:\ref{BB}, at each point and (-1,0) by grids. The gray region corresponds those enlarging lensing B-mode. The dots are their best fits.}
%\end{figure*}

%\begin{figure*}
%    \includegraphics[width=\columnwidth]{mcdes.png}
%    \caption{\label{mcdes}2D contours at 68\% and 95\% CL for
%        the parameters of the $w_0w_a$CDM model fitting to \textbf{Planck18+BK18+DESI+DESY5} dataset.}
%\end{figure*}

\section{Discussion}

In the concordant $\Lambda$CDM model, the simplest possibility of
DE is the cosmological constant. However, recently DESI DR2
combined with Planck CMB and supernova data has showed that DE is
evolving at $\gtrsim 3\sigma$ significance level. This result will
inevitably bring the shifts of the bestfit values of relevant
$\Lambda$CDM cosmological parameters. Here, using the latest
datasets we present the state-of-the-art constraints on the
tensor-to-scalar ratio $r$, which are tightened compared to the
result of BICEP/Keck collaboration.

%In corresponding $w_0w_a$CDM model, we find that the lower bounds
%of $r$ are slightly altered, while the bestfit $r$ is still
%$r_{0.05}\simeq 0.01$.

It is possible that relevant datasets have still some unknown
systematics, such as systematic uncertainties in B-mode
measurements, systematic errors in DESI DR2. It is also
interesting check how the result on $r$ changes under different
assumptions about the DE models (different parameterisations or
scalar fields models, e.g.recent \cite{Giare:2025pzu}),
reionization history, neutrino masses, or alternative priors.
Currently, it is still too early to say what about $r$, however,
our work is to not just say the search result for the primordial
GWs, but highlight how its detection is depending potentially on
models and datasets, i.e. our insight into the dark universe in
new era of cosmological surveys (DESI,
Euclid\cite{Euclid:2024yrr}, LSST
\cite{LSSTDarkEnergyScience:2018jkl}).

The scalar spectral index $n_s$ is the most crucial parameter
together with $r$ for understanding the physics of inflation. In
$\Lambda$CDM and $w_0w_a$CDM models, we have $n_s\simeq 0.96-0.97$
dependent of datasets, as seen in Table.\label{mctab}. However, it
is well known that both the concordant $\Lambda$CDM model and
evolving DE model suffered from the Hubble tension, see
e.g.\cite{Knox:2019rjx,Perivolaropoulos:2021jda,DiValentino:2021izs,Vagnozzi:2023nrq}.
In pre-recombination early dark energy (EDE) solution
\cite{Karwal:2016vyq,Poulin:2018cxd,Smith:2019ihp} to the Hubble
tension, in particular AdS-EDE solution
\cite{Ye:2020btb,Ye:2020oix,Jiang:2021bab,Wang:2022jpo},
%the upper bound on $r$ can be further tightened \cite{Ye:2022afu}.
$n_s=1$ since $n_s$ scales as ${\delta n_s}\simeq 0.4{\delta
H_0\over H_0}$ \cite{Ye:2021nej,Jiang:2022uyg,Ye:2022efx}. It has
been showed that such a pre-recombination EDE would also suppress
the shifts of $w_0$ and $w_a$ towards the evolving DE
\cite{Wang:2024dka}, see also \cite{Pang:2025lvh}, thus it is
significant to investigate the underlying impact of the Hubble
tension on not only the search for $r$ but the constraint on
$n_s$, e.g.\cite{Wang:2024tjd}.

\section*{Acknowledgments}

This work is supported by NSFC, No.12475064, National Key Research
and Development Program of China, No.2021YFC2203004, and the
Fundamental Research Funds for the Central Universities.

\end{document}